\documentclass[12pt,reqno]{amsart}
\usepackage{amsaddr}
\usepackage{amssymb,epsf}
\usepackage{amstext,amsmath,amsfonts,amscd,amsthm,graphicx}
\usepackage{upref}
\usepackage{epsfig}
\usepackage[utf8]{inputenc}
\usepackage{latexsym}
\usepackage{eucal}
\usepackage{wrapfig}
\usepackage{color}
\newfont{\fnt}{cmsy10}
\newfont{\sss}{cmti10}

\begin{document}

\title[Models of Rumour Impacts on Investment Decisions]{Information-Nonintensive Models of Rumour Impacts on Complex Investment Decisions}

\author{Nina~Bo\v{c}kov\'{a}$^1$, Karel~Doubravsk\'{y}$^2$, Barbora~Voln\'{a}$^3$*, Mirko~Dohnal$^2$}

\address{\footnotesize $^1$Prague University of Economics and Business, Faculty of Business Administration, W. Churchilla 4, \\ 130 67 Praha 3, Czech Republic, 
\\ $^2$Brno University of Technology, Faculty of Business and Management, Kolejn\'{i} 2906/4, \\ 61200 Brno, Czech Republic, 
\\ $^3$Silesian University in Opava, Mathematical Institute in Opava, Na Rybn\'{i}\v{c}ku 1, \\ 746 01 Opava, Czech Republic \\
*Corresponding author}
\email{nina.bockova@vse.cz}
\email{Karel.Doubravsky@vut.cz}
\email{Barbora.Volna@math.slu.cz}
\email{dohnal@vutbr.cz}


\begin{abstract}
This paper develops a qualitative framework for analysing the impact of rumours on complex investment decisions (CID) under severe information constraints. The proposed trend-based models rely on minimal data inputs in the form of increasing, decreasing, or constant relations. Sets of trend rules generate scenarios, and permitted transitions between them form a directed graph that represents system behaviour over time. The approach is applied in three interconnected models: financial CID, rumour-spreading dynamics, and their integration.
\end{abstract}

\maketitle

\section{Introduction}

Complex investment decisions (CID) are often unique, multidimensional, difficult to observe or measure, multi-objective, and multidisciplinary. They require the integration of several heterogeneous aspects, represented by both hard sciences (e.g., engineering) and soft sciences (e.g., finance, macroeconomics, sociology, and politics). A key problem affecting nearly all realistic CID-related forecasts is the shortage of information \cite{dohnal91,doubravsky_dohnal}. In fact, many important decision-making tasks suffer from severe shortages of input data and/or knowledge \cite{lee_kim_jung,skapa_bockova_doubravsky_dohnal}. Furthermore, some highly specific problems -- such as resilience-based decision making \cite{ardila-rueda_savachkin_romero-rodriguez_navarro} -- require the application of AI algorithms, again due to information scarcity. In addition, the predictive performance of traditional forecasting methods is also limited due to the inherent complexity of CID problems, see e.g., \cite{li_waheed_kirikkaleli_aziz}. Real-world investment processes are complex, integrated, ill-defined, and often difficult to observe -- particularly when, for example, ECG and macroeconomic parameters are taken into account \cite{skapa_bockova_doubravsky_dohnal}.

Experience, sparse or non-systematic industrial observations, and expert estimates represent a significant source of CID-related knowledge that cannot be formalised using conventional mathematical or statistical tools \cite{dwivedi_sharma_rana_giannakis_goel_dutot}. However, such vague knowledge, although difficult to formalise by traditional means, can be partially captured using alternative methods from artificial intelligence, including fuzzy and rough sets, genetic algorithms, fractal analysis, and qualitative reasoning \cite{dohnal83, dohnal_kocmanova, pavlakova-docekalova_doubravsky_dohnal_kocmanova}. This is especially relevant in cases where statistical methods -- relying on the law of large numbers -- fail due to data scarcity and problem complexity \cite{de-smith}. The primary sources of inherently subjective knowledge in CID include -- experience, analogy, feelings.

CIDs are intricate processes. Investors’ decisions can be significantly influenced by a broad spectrum of rumours of various types, originating from different companies or public influencers \cite{alzahrani_sarsam_al-samarraie_alblehai}. Fake financial news can cause temporary fluctuations in stock prices and increases in trading volume, resulting in abnormal trading behaviours \cite{cheng_lu_yeo}.

Rumours, which are widespread and historically persistent \cite{ding_xie}, tend to spread more rapidly when they contain negative or fear-inducing content, and they may significantly affect the economy \cite{ghosh_das_das22}. They are associated with a wide range of factors that are difficult to observe, measure, or predict, such as inferior product quality, pollution, financial fraud, and psychological responses like fear, uncertainty, and doubt \cite{alzahrani_sarsam_al-samarraie_alblehai}, as well as social and political aspects of events \cite{wang_vasilakos_ma_xiong}.

The traditional CID approach to human decision-making is characterised by attempts to optimise probability estimates and maximise expected utility. However, such models often fail, as the probability estimates are typically not sufficiently precise \cite{wang}.

Information technologies have significantly simplified the reproduction of information. An inevitable consequence is the increasing integration of CID and rumour models \cite{jahanbakhsh-nagadeh_feizi-derakhshi_ramezani}. This development creates substantial space for the spread of false information, such as fake news and rumours \cite{pattanaik_mandal_tripathy}. The propagation of rumours is becoming a significant social as well as macro- and microeconomic threat \cite{ghosh_das_das23}.

The presence of rumours -- typically unverified, ambiguous, and uncertain -- brings additional subjectivity into CID tasks. A real-world CID model integrated with a rumour component often suffers from severe shortages of information and knowledge. However, traditional statistical algorithms rely on extensive informational inputs. Trends -- i.e., increasing, decreasing, and constant values -- are among the least information-intensive quantifiers. A trend-based model can be constructed from a vague, common-sense, heterogeneous set of trend relations. Sets of trend rules are studied as generators of possible trend scenarios $S$. A set $T$ of transitions among the scenarios in $S$ is then used to produce unsteady-state behaviours, formalised by an oriented graph $H$. Any future or past behaviour of the CID-rumour system can be characterised by a path within the transitional graph $H$.

Our case study is based on three trend-based models:
\begin{enumerate}
\item Complex Investment Submodel (CIM) -- 7 scenarios; 10 variables, such as underpricing and return on assets,
\item Rumour-Related Submodel (RRM) -- 211 scenarios; 5 variables, such as the population of spreaders,
\item Integrated Model (IM) -- 14 scenarios; combining all 10 + 5 variables from CIM and RRM.
\end{enumerate}

Human-like common-sense reasoning appears useful for formalising vaguely known CID-related experience. Such reasoning often involves heuristics that are vaguely defined yet have been validated through long-standing tradition, see e.g., \cite{jung}:
\begin{quote}
\textit{No tree can grow to Heaven.}
\end{quote}
A simple reinterpretation of this heuristic is:
\begin{quote}
\textit{Quantitatively Unknown Upper Limit Exists.}
\end{quote}

A significant portion of CID data sets contains rumours or fake information, which seriously hinders the further processing of extracted knowledge or data items \cite{fu_feng_lande_dmytrenko_manko_prakapovich, huo_chen_zhao, stieglitz_mirbabaie_ross_neuberger}. As a result, problems related to rumour propagation are increasingly recognised as important, see e.g., \cite{cheng_lu_yeo, wang_song}.

In summary, this paper provides a qualitative framework for integrating rumour dynamics into CID modelling under severe information constraints, using scenario-based trend reasoning and transition graphs to formalise and simulate possible system behaviours.

\section{Formal Models of Rumour}

Classical models of rumour dynamics are typically based on three groups of participants; see e.g. \cite{sun_sheng_cui}:
\begin{itemize}
\item \textit{Ignorants} -- individuals who are not aware of the rumour,
\item \textit{Spreaders} -- individuals who actively spread rumour-related information,
\item \textit{Stiflers} -- those who have heard the rumour but choose not to propagate it.
\end{itemize}

A persistent challenge in CID forecasting is the shortage of reliable information \linebreak \cite{dohnal91,doubravsky_dohnal}. Combined with the inherent complexity of CID systems, this significantly limits the effectiveness of traditional forecasting methods \cite{li_waheed_kirikkaleli_aziz}. As a result, formal approaches such as fuzzy and rough sets are increasingly employed to address these limitations \cite{dohnal85,zhang_xie_wang}.

Partial mitigation of severe information shortages can be achieved by incorporating non-numerical knowledge into rumour analysis. These methods must account for relevant information and knowledge that may be embedded in available (semi-) subjective inputs \cite{skapa_bockova_doubravsky_dohnal}.

The negative effects of small CID data sets and poor-quality information or knowledge can also be addressed by non-statistical grey forecasting algorithms \cite{liu_yang_forrest}. Grey forecasting methods distinguish between known, partially known, and unknown information items \cite{jalali_heidari,liu_yang_forrest}.

A well-understood set of qualitative relations among variables can form a meaningful model when traditional quantitative approaches require an unacceptably large amount of data or knowledge to be effective. Qualitative or trend-based models can assist in identifying variables that play a key role in shaping relevant relationships, particularly in data-poor contexts \cite{mancini_coghill_lusseau}.

One type of common-sense reasoning that has been particularly successful is known as qualitative reasoning \cite{davis_marcus}. This theory has been applied across many domains, from management to engineering \cite{de-kleer_MIT,de-kleer_brown}. However, the term \textit{qualitative} has multiple interpretations -- e.g., in the context of qualitative data sets -- which can lead to confusion. For this reason, the term \textit{trend} is used in this paper instead.

\section{Trend Models}

Trend-reasoning research develops non-numerical reasoning techniques \cite{de_kleer}. It enables systems to reason about behaviour without requiring the precise quantitative information typically needed by conventional analysis methods, such as numerical simulators \cite{nwaibeha_chikwendub}. Compared to traditional numerical simulation, trend reasoning offers several advantages, including the ability to cope with incomplete information, provide imprecise but valid predictions, and facilitate easier interpretation of results \cite{bobrow,kuipers,zaslow}. Building trend models involves transforming vague and general ideas into more formalised structures \cite{bredeweg_salles_bouwer}, and can support a broad spectrum of knowledge-based tasks \cite{bredeweg,saeed_omlin}.

The formal tool is presented in this paper for rumour analysis, based on \textit{trend quantities}: positive, zero, negative, or unknown, associated with model variables. The use of trend quantities is presented in Table~\ref{tab:trend_quantifiers}, where the first row lists the corresponding symbols, the second expresses their qualitative meaning with respect to variable values, and the third expresses their meaning in terms of first-order trends -- that is, the qualitative interpretation of first derivatives (increasing, constant, decreasing).
\begin{table}[ht]
\centering
\begin{tabular}{l l l l l}
\hline
\textbf{Symbol}           & $+$        & $0$      & $-$          & $*$            \\
\hline
\textbf{Value}            & Positive   & Zero     & Negative     & Arbitrary      \\
\textbf{Trend derivative} & Increasing & Constant & Decreasing   & Any direction  \\
\hline \hline
\end{tabular}
\vspace{0.2cm}
\caption{Trend quantifiers}
\label{tab:trend_quantifiers}
\end{table}
These trend quantifiers provide a concise way to represent qualitative information about the basic behaviour of a studied variable. 

In addition, trend-based analysis can describe pairwise shallow knowledge relations between model variables $X$ and $Y$. The following types of such relations are used to formalise qualitative information statements:
\begin{equation}
\label{SUP_RED}
\begin{array}{l}
\text{An increase in $X$ has a \textit{supporting} effect on $Y$ (labelled SUP$XY$).} \\
\text{An increase in $X$ has a \textit{reducing} effect on $Y$ (labelled RED$XY$).}
\end{array}
\end{equation}

The following set of relations \cite{sachdeva_lehal} illustrates the flexibility of trend-based analysis grounded in \eqref{SUP_RED}:
\begin{itemize}
\item Personal financial needs positively influence investment decision making.
\item Firm image positively influences investment decision making.
\item Neutral information positively influences investment decision making.
\end{itemize}

\begin{figure}[ht]
  \centering
  \includegraphics[height=6cm]{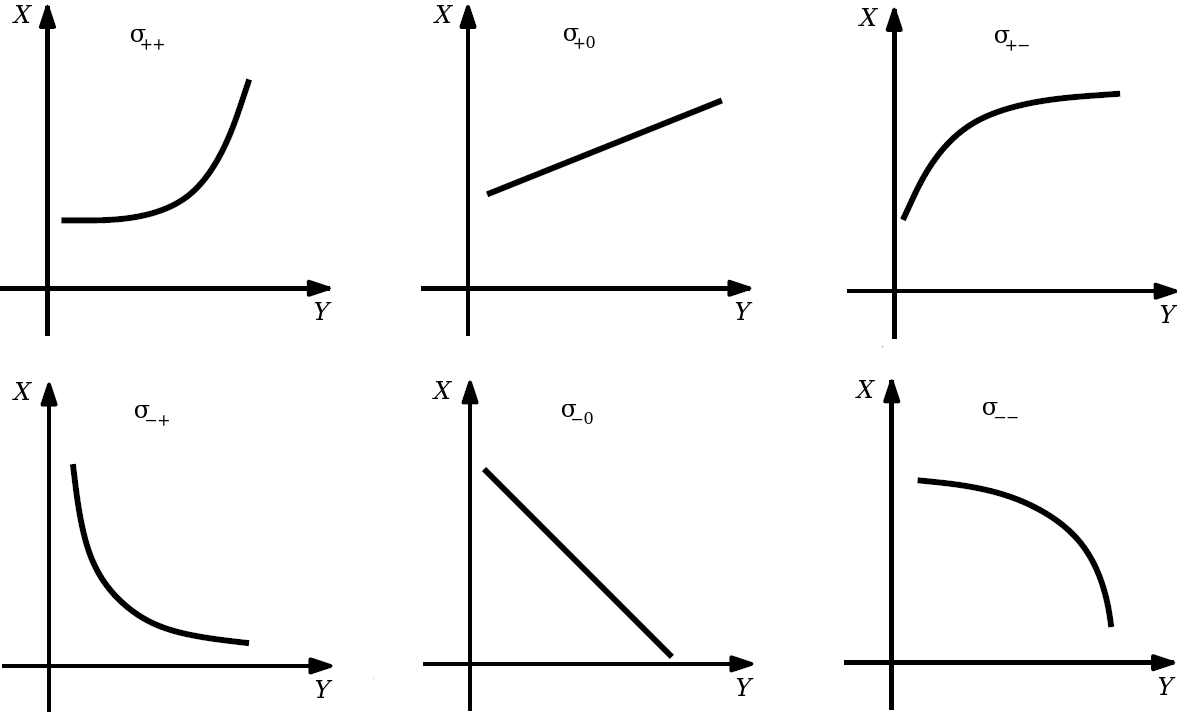}
  \caption{Graphical representation of qualitative shapes $\sigma$ illustrating possible relationships between variables $X$ and $Y$ based on the first and second trend derivatives ($D$, $DD$).}
  \label{fig:shapes}
\end{figure}

The trend derivatives -- see Table~\ref{tab:trend_quantifiers} -- are a qualitative analogue of numerical first derivatives. However, some information or knowledge can be formalised with higher precision. In such cases, second-order trends, i.e. trends of trends, are used as an analogue of second derivatives. Let $X$ and $Y$ be studied variables, and let their values be assumed positive. Then, based on the first and second trend derivatives, denoted by $D$ and $DD$ respectively, there are six possible shapes $\sigma$ illustrating the qualitative relationship between $X$ and $Y$. These shapes are represented by the graphs of functions $Y(X)$ in Figure~\ref{fig:shapes}. The first and second subscripts indicate the signs of the first and second derivatives, respectively, of the function $Y(X)$. For example, $\sigma_{+-}$ represents a case where $D$ is positive and $DD$ is negative.
The trend shapes $\sigma_{++}$, $\sigma_{+0}$, and $\sigma_{+-}$ represent supporting effects, while the shapes $\sigma_{-+}$, $\sigma_{-0}$, and $\sigma_{--}$ represent reducing effects; see equation \eqref{SUP_RED}.

We now define the concept of a \textit{trend model}. Let $X_i$, for $i = 1, 2, \ldots, n$, denote the set of model variables. Then, the trend model, denoted by $M$, is defined as a collection of pairwise trend relations:
\begin{equation}
\label{pairwise_relations}
P_k(X_i, X_j), \quad k = 1, 2, \ldots, w,
\end{equation}
where $X_i \neq X_j$. Thus, the trend model is given by:
\begin{equation}
\label{trend_model}
M = \left\{ P_k(X_i, X_j) \;\middle|\; k = 1, 2, \ldots, w; \; i, j = 1, 2, \ldots, n; \; i \neq j \right\}.
\end{equation}
The supporting and reducing effects \eqref{SUP_RED} are used to describe these pairwise relations \eqref{pairwise_relations}. Moreover, in cases where the second derivatives $DD$ are known, a more refined qualitative description based on shape categories can be applied; see Figure~\ref{fig:shapes}. Unfortunately, in many rumour-related tasks with limited information, only the first derivatives are available. In such cases, the supporting and reducing effects \eqref{SUP_RED} provide sufficient expressive power for constructing the trend model.

The main output of trend-based modelling is the set of possible scenarios and possible transitions between them. This set of scenarios and transitions can be seen as the solution to the trend model. Finding this solution represents a combinatorial task and is not studied in this paper; for details, see \cite{dohnal_doubravsky15}. There are various interpretations of different types of scenarios; see e.g. \cite{cliff_noss_anderson}. The concept of a \textit{scenario}, as used in this paper, is defined by the following set of $n$ trend triplets:
\begin{equation}
\label{scenarios}
S(n,m) = \left\{ \{(X_1 DX_1 DDX_1), (X_2 DX_2 DDX_2), \ldots, (X_n DX_n DDX_n)\}_j \right\},
\end{equation}
where $j = 1, 2, \ldots, m$, $DX_i$ is the first and $DDX_i$ the second time trend derivative of the $i$-th variable, expressed using trend quantities $+$, $0$, or $-$; see Table~\ref{tab:trend_quantifiers}. The set of scenarios \eqref{scenarios} contains $m$ scenarios and is $n$-dimensional in the sense that it refers to $n$ model variables. We can identify the following particular scenario \eqref{steady_state} from the set $S(n,m)$ as a steady state:
\begin{equation}
\label{steady_state}
\{(X_1, 0, 0), (X_2, 0, 0), \ldots, (X_n, 0, 0)\}
\end{equation}
In this scenario, both the first and second trend derivatives are zero for all model variables, analogous to the steady-state condition in the quantitative case.

The graphical representation of possible transitions between scenarios from the set $S(n, m)$ \eqref{scenarios} is referred to as the \textit{transitional graph}. A transitional graph, denoted by $H$, is a directed graph with scenarios from the set $S$ \eqref{scenarios} as nodes and transitions from the set $T$ as directed edges:
\begin{equation}
\label{directed_graph}
H = (S,T)
\end{equation}
The elements of the set $T$ represent transitions between scenarios. These transitions are based on rules specified for an individual model variable with a positive value (i.e. $+$ in the first position of the corresponding triplet), which remains positive after the transition, as summarised in Table~\ref{tab:transitions}; for details, see \cite{dohnal_kocmanova}. These transitions are then applied analogously to all model variables.
\begin{table}[ht]
\centering
\begin{tabular}{c l l l l}
\hline
No. & From & To & Alt. 1 & Alt. 2  \\						
1   & $+++$ & $++\,0$ & & \\						
2   & $++\,0$ & $+++$ & $++-$ & \\
3   & $++-$ & $++\,0$ & $+~0~-$ & $+~0~0$ \\				
4   & $+~0~+$ & $+++$ & & \\						
5   & $+~0~0$ & $+++$ & $+--$ & \\					
6   & $+~0~-$ & $+--$ & & \\						
7   & $+-+$ & $+-\,0$ & $+~0~+$ & $+~0~0$ \\
8   & $+-\,0$ & $+-+$ & $+--$ &   \\				
9   & $+--$ & $+-\,0$ & & \\				
\hline \hline
\end{tabular}
\vspace{0.2cm}
\caption{Transition rules for a positively valued variable}
\label{tab:transitions}
\end{table}
For example, the second line of Table~\ref{tab:transitions} indicates that it is possible to transition from the triplet $(++0)$ to either $(+++)$ or $(++-)$. These rules describe the time behaviour of the model variable. Such changes are expected to be continuous and smooth, without any jumps, steps, or kinks. In mathematical terms, this means that the corresponding time function is continuous, smooth, or at least continuously differentiable.

In traditional (numerical) settings, the well-known correlation matrix $C$ \eqref{correlation_matrix} is used to describe pairwise relations between all model variables:
\begin{equation}
\label{correlation_matrix}
C = \left(
\begin{array}{c c c}
c_{11} & \dots & c_{1n} \\
\vdots & \ddots & \vdots \\
c_{n1} & \dots & c_{nn}
\end{array}
\right),
\end{equation}
where $c_{ij}$, for $i,j = 1, \ldots, n$, are correlation coefficients between model variables. In the trend-based approach, positive and negative values of $c_{ij}$ can be interpreted as supporting (SUP${XY}$) and reducing (RED${XY}$) effects between variables $X$ and $Y$, respectively.

Unfortunately, the collection of pairwise trend relations $P_k(X_i, X_j)$, $k = 1, 2, \dots, w$ \eqref{pairwise_relations}, derived from the correlation matrix \eqref{correlation_matrix} using supporting and reducing effects, often leads to meaningless solutions of the trend model -- for instance, a scenario set containing only the steady state scenario \eqref{steady_state}. Such cases reflect situations where the resulting set of pairwise trend relations contains a contradictory subset -- for example, when a variable is both increasing and decreasing during the same time period. In such cases, we say that the set of pairwise trend relations is inconsistent. These inconsistencies should be removed, and various removal algorithms exist for this purpose. The simple heuristic algorithm used in this paper proceeds as follows:
\begin{quote}
\itshape
Remove the correlation coefficient $c_{ij}$ with the smallest absolute value from the correlation matrix~\eqref{correlation_matrix}. Then test the trend model derived from the updated correlation matrix. If the resulting trend solution is still the steady-state scenario~\eqref{steady_state}, repeat this heuristic.
\end{quote}

\newpage
\section{Case Study}

In this section, we present a trend-based Complex Investment Model (CIM) and a trend-based Rumour-Related Model (RRM) as preparatory components, followed by their integration into an aggregate trend-based Integrated Model (IM).

\subsection{Complex Investment Submodel (CIM)}

The original CIM was published in \cite{salerno_sampagnaro_verdoliva} in the form of a correlation matrix. The variables used in the CIM are listed in Table~\ref{tab:CIM_variables}.
\begin{table}[ht]
\centering
\begin{tabular}{c l}
\hline
\textbf{Abbreviation} & \textbf{Variable} \\
\hline
UND & Underpricing \\
AGE & Age \\
TA  & Size \\
MAR & Market Capitalization \\
LIS & Number of Listed Domestic Companies \\
QUA & Underwriter Syndicate \\
REP & Number of IPOs Underwritten by Syndicates \\
BOO & Book-to-Market Ratio \\
ROA & Return on Assets \\
PRI & Price-to-Book Ratio \\
\hline \hline
\end{tabular}
\vspace{0.2cm}
\caption{Variables in the Complex Investment Model (CIM)}
\label{tab:CIM_variables}
\end{table}

The collection of pairwise trend relations derived from this original matrix suffers from the inconsistencies discussed earlier. Therefore, it is processed using the removal algorithm described above and modified by incorporating (semi-)subjective information -- such as experience, analogy, and feelings -- provided by a team of experts. The result of this procedure is a set of 14 pairwise trend relations forming the trend-based CIM; see Table~\ref{tab:CIM}.
\begin{table}[ht]
\centering
\begin{tabular}{c l}
\hline
No. & Model Row \\
\hline
1  & RED UND TA \\
2  & RED AGE ROA \\
3  & SUP QUA TA \\
4  & SUP MAR LIS \\
5  & SUP MAR REP \\
6  & SUP LIS QUA \\
7  & SUP LIS REP \\
8  & $\sigma_{+-}$ REP PRI \\
9  & SUP QUA PRI \\
10 & SUP BOO TA \\
11 & SUP BOO LIS \\
12 & SUP AGE PRI \\
13 & SUP LIS REP \\
14 & SUP QUA REP \\
\hline \hline
\end{tabular}
\vspace{0.2cm}
\caption{Pairwise trend relations defining the CIM}
\label{tab:CIM}
\end{table}
Thus, Table~\ref{tab:CIM} lists the resulting trend relations. The first column contains row numbers, and the second column shows the corresponding pairwise trend relations based on \eqref{SUP_RED}, derived from the modified correlation matrix or, in one case, described more precisely using qualitative trend shapes $\sigma$ as illustrated in Figure~\ref{fig:shapes}.

The trend-based CIM, see Table \ref{tab:CIM}, has 7 scenarios as its solution listed in Table \ref{tab:CIM_scenarios}.
\begin{table}[ht]
\centering
\begin{tabular}{c l l l l l l l l l l}
\hline
No. & REP & UND & AGE & TA & MAR & LIS & QUA & BOO & ROA & PRI \\
\hline
1 & $+++$ & $+--$ & $+++$ & $+++$ & $+++$ & $+++$ & $+++$ & $+++$ & $+--$ & $+++$ \\
2 & $++-$ & $+-+$ & $++-$ & $++-$ & $++-$ & $++-$ & $++-$ & $++-$ & $+-+$ & $++-$ \\
3 & $+~0~+$ & $+~0~-$ & $+~0~+$ & $+~0~+$ & $+~0~+$ & $+~0~+$ & $+~0~+$ & $+~0~+$ & $+~0~-$ & $+~0~+$ \\
4 & $+~0~0$ & $+~0~0$ & $+~0~0$ & $+~0~0$ & $+~0~0$ & $+~0~0$ & $+~0~0$ & $+~0~0$ & $+~0~0$ & $+~0~0$ \\
5 & $+~0~-$ & $+~0~+$ & $+~0~-$ & $+~0~-$ & $+~0~-$ & $+~0~-$ & $+~0~-$ & $+~0~-$ & $+~0~+$ & $+~0~-$ \\
6 & $+-+$ & $++-$ & $+-+$ & $+-+$ & $+-+$ & $+-+$ & $+-+$ & $+-+$ & $++-$ & $+-+$ \\
7 & $+--$ & $+++$ & $+--$ & $+--$ & $+--$ & $+--$ & $+--$ & $+--$ & $+++$ & $+--$ \\
\hline \hline
\end{tabular}
\vspace{0.2cm}
\caption{Scenarios of the trend-based CIM}
\label{tab:CIM_scenarios}
\end{table}
It is easy to see that scenario no. 4 corresponds to a steady state. Moreover, we can observe an interesting phenomenon: the variables REP, AGE, TA, MAR, LIS, QUA, BOO, and PRI share the same set of scenarios, as do the variables UND and ROA. This indicates that the variables in each group exhibit identical trend behaviour over time.

\subsection{Rumour-Related Submodel (RRM)}

The autonomous system of first-order differential equations given in \eqref{system_RRM}, adapted from \cite{chen_chen_song}, is used as a model of rumour spreading.
\begin{equation}
\label{system_RRM}
\begin{array}{r c l}
\frac{dX}{dt}   &=& -\alpha \cdot \frac{X \cdot Y}{N} \\
\frac{dY}{dt}   &=& \varphi \cdot \alpha \cdot \frac{X \cdot Y}{N} - \delta \cdot \frac{Y \cdot (Y + Z_1)}{N}-\lambda \cdot \frac{Y \cdot Z_2}{N}\\
\frac{dW}{dt}   &=& (1-\varphi) \cdot \alpha \cdot \frac{X \cdot Y}{N}-\eta \cdot W \\
\frac{dZ_1}{dt} &=& \delta \cdot \frac{Y \cdot (Y + Z_1)}{N} + \theta \cdot \eta \cdot W \\
\frac{dZ_2}{dt} &=& (1-\theta) \cdot \eta \cdot W + \lambda \cdot \frac{Y \cdot Z_2}{N} \\
\end{array}
\end{equation}

In this system, there are five variables, listed in Table~\ref{tab:RRM_variables}, and six unknown positive parameters $\alpha, \varphi, \delta, \lambda, \theta$, and $\eta$. The constant $N > 0$ denotes the total population.
\begin{table}[ht]
\centering
\begin{tabular}{c l}
\hline
\textbf{Label} & \textbf{Variable} \\
\hline
$X$   & Population of ignorants \\
$Y$   & Population of spreaders \\
$W$   & Population of sceptics \\
$Z_1$ & Number of promoters who support the negative rumour \\
$Z_2$ & Number of suppressors who oppose the negative rumour \\
\hline \hline
\end{tabular}
\vspace{0.2cm}
\caption{Variables in the Rumour-Related Model (RRM)}
\label{tab:RRM_variables}
\end{table}

The original system of differential equations contains seven numerical constants. Their traditional identification is problematic due to the typical severe information shortages; see e.g. \cite{verdult_verhaegen}. However, when using trend quantifiers, numerical multiplicative constants such as $A > 0$ or $B < 0$ can generally be eliminated \cite{dohnal16, doubravsky_dohnal}:
\begin{equation}
\label{constant_trend}
\begin{array}{l}
AX = (+)X = X \\
BX = (-)X = -X
\end{array}
\end{equation}
Moreover, the following simple expressions are used to apply trend-based analysis to the system of differential equations; for details, see e.g. \cite{forbus,konecny_vicha_dohnal}:
\begin{equation}
\label{addition_trend}
\begin{array}{l}
(+) + (+) = (+) \\
(+) + (-) = (+ \text{ or } 0 \text{ or }-)
\end{array}
\end{equation}

The system of numerically based differential equations~\eqref{system_RRM} is translated into the trend-based RRM~\eqref{trend_RRM} using the conversion rules given in \eqref{constant_trend} and \eqref{addition_trend}.
\begin{equation}
\label{trend_RRM}
\begin{array}{r c l}
DX + XY               &=& 0 \\
DY + YY + YZ_1 + YZ_2 &=& XY\\
DW + XY + W           &=& XY \\
DZ_1                  &=& YY + YZ_1 + W \\
DZ_2 + W              &=& W + YZ_2 \\
\end{array}
\end{equation}

The combinatorial algorithm used to generate the scenario set is not specified in this paper; for details, see e.g. \cite{dohnal_doubravsky15, dohnal_doubravsky16}. The trend-based RRM~\eqref{trend_RRM} yields 211 scenarios. To maintain readability, the full scenario set is not provided here. This submodel serves as a preparatory component for the construction of the aggregate IM.

\subsection{Trend-based Integrated Model (IM)}

The models CIM (Table~\ref{tab:CIM}) and RRM (Equation~\eqref{trend_RRM}) are integrated into the IM via the union:
\begin{equation}
\label{IM}
\text{IM} \equiv \text{CIM} \cup \text{RRM}
\end{equation}
and further extended by three additional pairwise trend relations listed in Table~\ref{tab:IM_additional}. These relations refer to a subset of the variables listed in Tables~\ref{tab:CIM_variables} and~\ref{tab:RRM_variables}, and are based on common-sense reasoning, the supporting and reducing effects \eqref{SUP_RED}, and the qualitative trend shapes $\sigma$ (Figure~\ref{fig:shapes}).
\begin{table}[ht]
\centering
\begin{tabular}{l}
\hline
Model Row \\
\hline
$\sigma_{+-}$ Z2 REP \\
$\sigma_{--}$ Z1 UND \\
RED W REP \\
\hline \hline
\end{tabular}
\vspace{0.2cm}
\caption{Additional pairwise trend relations included in the IM}
\label{tab:IM_additional}
\end{table}

The trend-based IM, defined by Equation~\eqref{IM} and the additional relations in Table~\ref{tab:IM_additional}, yields 14 scenarios, which are provided in Table~\ref{tab:IM_scenarios} as its solutions.
\begin{table}[ht]
\centering
\begin{tabular}{c l l l l l l l}
\hline
No. & REP & ROA & $X$ & $Y$ & $W$ & $Z_1$ & $Z_2$ \\
\hline
1  & $+++$ & $+--$ & $+-+$ & $+-+$   & $+--$ & $++-$   & $+++$ \\
2  & $+++$ & $+--$ & $+-+$ & $+-\,0$ & $+--$ & $++-$   & $+++$ \\
3  & $+++$ & $+--$ & $+-+$ & $+--$   & $+--$ & $+++$   & $+++$ \\
4  & $+++$ & $+--$ & $+-+$ & $+--$   & $+--$ & $++\,0$ & $+++$ \\
5  & $+++$ & $+--$ & $+-+$ & $+--$   & $+--$ & $++-$   & $+++$ \\
6  & $++-$ & $+-+$ & $+-+$ & $+-+$   & $+-+$ & $++-$   & $+++$ \\
7  & $++-$ & $+-+$ & $+-+$ & $+-+$   & $+-+$ & $++-$   & $++\,0$ \\
8  & $++-$ & $+-+$ & $+-+$ & $+-+$   & $+-+$ & $++-$   & $++-$ \\
9  & $++-$ & $+-+$ & $+-+$ & $+-\,0$ & $+-+$ & $++-$   & $+++$ \\
10 & $++-$ & $+-+$ & $+-+$ & $+-\,0$ & $+-+$ & $++-$   & $++\,0$ \\
11 & $++-$ & $+-+$ & $+-+$ & $+-\,0$ & $+-+$ & $++-$   & $++-$ \\
12 & $++-$ & $+-+$ & $+-+$ & $+--$   & $+-+$ & $++-$   & $+++$ \\
13 & $++-$ & $+-+$ & $+-+$ & $+--$   & $+-+$ & $++-$   & $++\,0$ \\
14 & $++-$ & $+-+$ & $+-+$ & $+--$   & $+-+$ & $++-$   & $++-$ \\
\hline \hline
\end{tabular}
\vspace{0.2cm}
\caption{Scenarios of the trend-based IM}
\label{tab:IM_scenarios}
\end{table}
Consistent with the solution of the CIM, the variables REP, AGE, TA, MAR, LIS, QUA, BOO, and PRI share identical trend triplets across all 14 scenarios; similarly, the variables UND and ROA do as well. Therefore, only one representative from each subset is included in Table~\ref{tab:IM_scenarios}, specifically REP and ROA.

If the second derivative $DD$ in \eqref{scenarios} is not taken into consideration for any model variable, the integrated model IM~\eqref{IM} together with the additional relations in Table~\ref{tab:IM_additional} yields only a single scenario, shown in Table~\ref{tab:IM_one_scenario}.
\begin{table}[ht]
\centering
\begin{tabular}{l l l l l l l}
\hline
REP & ROA & $X$ & $Y$ & $W$ & $Z_1$ & $Z_2$ \\
\hline
$++*$ & $+-*$ & $+-*$ & $+-*$ & $+-*$ & $++*$ & $++*$ \\
\hline \hline
\end{tabular}
\vspace{0.2cm}
\caption{Scenario of the trend-based IM without second derivatives}
\label{tab:IM_one_scenario}
\end{table}
This scenario represents a situation in which the variables REP, $Z_1$, and $Z_2$ are always increasing, while the variables ROA, $X$, $Y$, and $W$ are always decreasing. For the interpretation of trend quantifiers, see Table~\ref{tab:trend_quantifiers}.

\begin{figure}[ht]
  \centering
  \includegraphics[height=6cm]{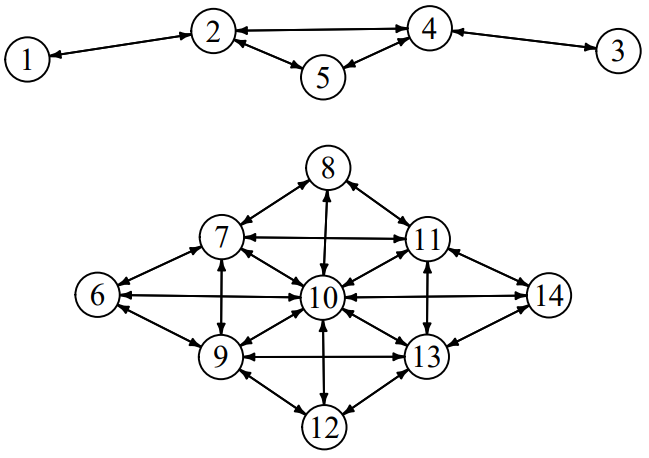}
  \caption{Transitional graph based on the scenario set of the trend-based IM}
  \label{fig:IM_transitional_graph}
\end{figure}

However, when second derivatives $DD$ are taken into consideration, the solution of the IM exhibits unsteady-state behaviour; see Table~\ref{tab:IM_scenarios}. The corresponding transitional graph, based on this set of 14 scenarios, is illustrated in Figure~\ref{fig:IM_transitional_graph}.

Multi-Criteria Decision Analysis (MCDA), see e.g. \cite{jamwa_agrawal_sharma_kumar}, helps decision-makers avoid poor investment choices when multiple factors must be considered. This is particularly relevant today, as decision-making procedures have become increasingly complex with the involvement of more investors -- each bringing a specific and often partially conflicting set of priorities \cite{abdul-kareem_fayed_rady_el-regaily_nema,jain_walia_gupta}.

Let us focus on three variables: REP, ROA, and UND. Due to their nature, all three are assumed to be maximisation goals. The best possible trend behaviour for a variable is described by the triplet $(+++)$, representing steep accelerating growth. However, no scenario in Table~\ref{tab:IM_scenarios} includes this triplet for all three selected variables simultaneously. It follows that a compromise is inevitable. Scenarios Nos. 1--5 contain the triplet $(+++)$ for REP. However, in these scenarios, ROA and UND are described by the worst maximisation triplet $(+--)$, which corresponds to steep accelerating decline. If an investor chooses to ignore the non-optimal behaviour of ROA and UND, then scenarios 1--5 represent the best available options for maximising REP. Importantly, this subset of scenarios cannot be reached from scenarios 6--14, as shown in the transitional graph (Figure~\ref{fig:IM_transitional_graph}). Conversely, if an investor prefers gradual changes over time, scenarios 6--14 offer the triplet $(++-)$ for REP -- representing gradual slowing growth -- and the triplet $(+-+)$ for ROA and UND, corresponding to a gradual slowing decline.

\section{Conclusion}

Trend models can be viewed as tools that support and complement human decision-making rather than replace it. Combining human intuition with algorithmic assistance allows us to leverage the strengths of trend-based analysis while preserving accountability and control over decisions.

Financial institutions are increasingly adopting trend models to support investors in making more informed choices. These models help reduce cognitive workload, allowing investors to focus on tasks that require human expertise -- such as monitoring market developments or evaluating the impact of geopolitical events. Integrating trend models into the decision-making process enables portfolio optimisation by identifying strategies that align with specific goals while reducing risk. In many cases, this approach contributes to the potential for improved investment performance.

Artificial intelligence research has long examined how decision-makers and system designers generate explanations and make context-sensitive judgments; see e.g. \cite{davis_marcus,dohnal85,dohnal91,dohnal16,dwivedi_sharma_rana_giannakis_goel_dutot,forbus,li_waheed_kirikkaleli_aziz,skapa_bockova_doubravsky_dohnal}. Trend models are a result of these efforts. They offer a formalised framework for representing human-like reasoning in dynamic domains, where variables evolve through qualitative state changes.

\newpage
The key contributions of this paper are:
\begin{itemize}
\item We apply a recent approach based on common-sense trend analysis.
\item We demonstrate that the effectiveness of decision-makers who are neither mathematicians nor experts in artificial intelligence can be increased through the use of trend-based analysis.
\item We confirm the ongoing effort among theoreticians and practitioners to develop transparent, easy-to-understand AI models.
\end{itemize}

The main advantages of the proposed trend-based method include:
\begin{itemize}
\item No numerical values of constants or parameters are required, and the resulting set of trend solutions forms a superset of all meaningful solutions.
\item A complete list of possible futures and histories can be generated.
\item The results are easy to interpret without requiring advanced mathematical knowledge.
\end{itemize}

Humans are naturally able to understand their environment and interact with objects and processes that undergo dynamic transformations, making at least approximate predictions about how observed events will unfold. Trend models attempt to reflect this type of reasoning within structured decision-support tools.

\section*{Acknowledgements}
The research was supported by Prague University of Economics and Business (Faculty of Business Administration), Brno University of Technology (Faculty of Business and Management) and Silesian University in Opava (Mathematical Institute in Opava), Czech Republic.

\section*{Statement}
During the preparation of this work the authors used ChatGPT (OpenAI) in order to improve English. After using this tool, the authors reviewed and edited the content as needed and take full responsibility for the content of the publication.


\begin{thebibliography}{99}
\bibitem{abdul-kareem_fayed_rady_el-regaily_nema} A.A. Abdul Kareem, Z.T. Fayed, S. Rady, S. A. El-Regaily, B.M. Nema, 2023. Factors Influencing Investment Decisions in Financial Investment Companies. Systems 11, 146. https://doi.org/10.3390/systems11030146.
\bibitem{ardila-rueda_savachkin_romero-rodriguez_navarro} W. Ardila-Rueda, A. Savachkin, D. Romero-Rodriguez, J. Navarro, 2025. Balancing the costs and benefits of resilience-based decision making. Decis. Support Syst. 191, 114425. https://doi.org/10.1016/j.dss.2025.11442.
\bibitem{alzahrani_sarsam_al-samarraie_alblehai} A. I. Alzahrani, S. M. Sarsam, H. Al-Samarraie, F. Alblehai, Exploring the sentimental features of rumor messages and investors’ intentions to invest, Internat. Rev. Econom. Finance 87 (2023) 433--444. https://doi.org/10.1016/j.iref.2023.05.006.
\bibitem{bredeweg} B. Bredeweg, Modelling Problem Solving (Chapter 3), in: Expertise in Qualitative Prediction of Behaviour, PhD thesis, University of Amsterdam, 1992.
\bibitem{bredeweg_salles_bouwer} B. Bredeweg, P. Salles, A. Bouwer et al., Towards a structured approach to building qualitative reasoning models and simulations, Ecol. Inform. 3 (1) (2008) 1--12. https://doi.org/10.1016/j.ecoinf.2007.02.002.
\bibitem{bobrow} D.G. Bobrow, Qualitative reasoning about physical systems, first ed., Elsevier, North Holland, 1985.
\bibitem{chen_chen_song} J. Chen, C. Chen, Q. Song et al., 2021. Spread Mechanism and Control Strategies of Rumor Propagation Model Considering Rumor Refutation and Information Feedback in Emergency Management. Symmetry 13 (9), 1694. https://doi.org/10.3390/sym13091694.
\bibitem{cheng_lu_yeo} L.-C. Cheng,  W.-T. Lu, B. Yeo, 2023. Predicting abnormal trading behavior from internet rumor propagation: a machine learning approach. Financ. Innov. 9, 3. https://doi.org/10.1186/s40854-022-00423-9.
\bibitem{cliff_noss_anderson} P. Cliff, J. Noss, G. Anderson at al., Scenario Analysis Guide for Asset Managers, Climate Financial Risk Forum Guide 2023, FCA Public, 2023.
\bibitem{davis_marcus} E. Davis, G. Marcus, Commonsense reasoning and commonsense knowledge in artificial intelligence, Commun. ACM 58 (9) (2915) 92--103. https://doi.org/10.1145/2701413.
\bibitem{de_kleer} J. de Kleer, Structural Knowledge in Troubleshooting Circuits, in: J.S. Brown (Ed), Steps toward a theoretical foundation for complex knowledge-based CAI, Bolt Beranek and Newman, Cambridge, Massachusetts, 1975, pp. 52-78.
\bibitem{de-kleer_MIT} J. de Kleer, Qualitative and quantitative knowledge in classical mechanics, MIT AI Lab, 1975.
\bibitem{de-kleer_brown} J. De Kleer, J. S. Brown, A qualitative physics based on confluences, Artificial Intelligence 24 (1-3) (1984) 7--83. https://doi.org/10.1016/0004-3702(84)90037-7.
\bibitem{de-smith} M. J. De Smith, Statistical Analysis Handbook, hardback ed., Winchelsea Press, UK, 2018.
\bibitem{ding_xie} H. Ding, L. Xie, 2023. Simulating rumor spreading and rebuttal strategy with rebuttal forgetting: An agent-based modeling approach. Phys. A 612, 128488. https://doi.org/10.1016/j.physa.2023.128488.
\bibitem{dohnal83} M. Dohnal, Fuzzy simulation of industrial problems, Comput. Indust. 4 (4) (1983) 347--352. https://doi.org/10.1016/0166-3615(83)90002-7.
\bibitem{dohnal85} M. Dohnal, Applications of a universal expert system in industry. Comput. Indust. 6 (2) (1985) 115--121. http://dx.doi.org/10.1016/0166-3615(85)90017-X.
\bibitem{dohnal91} M. Dohnal, A methodology for common-sense model development, Comput. Indust. 16 (2) (1991) 141--158. https://doi.org/10.1016/0166-3615(91)90086-O.
\bibitem{dohnal16} M. Dohnal, Complex bio fuels related scenarios generated by qualitative reasoning under severe information shortages: A review. Renewable Sustainable Energy Rev. 65 (2016) 676--684. https://doi.org/10.1016/j.rser.2016.07.029.
\bibitem{dohnal_doubravsky15} M. Dohnal, K. Doubravsk\'{y}, 2015. Qualitative Upper and Lower Approximations of Complex Nonlinear Chaotic and Nonchaotic Models. Internat. J. Bifur. Chaos. Appl. Sci Engrg. 25 (13) 1550173. https://doi.org/10.1142/S0218127415501734.
\bibitem{dohnal_doubravsky16} M. Dohnal, K. Doubravsk\'{y}, Equationless and equation-based trend models of prohibitively complex technological and related forecasts. Tech. Forecast. Soc. Change 111 (2016) 297--304. https://doi.org/10.1016/j.techfore.2016.07.031. 
\bibitem{dohnal_kocmanova} M. Dohnal, A. Kocmanov\'{a}, Qualitative models of complex sustainability systems using integrations of equations and equationless knowledge items generated by several experts. Ecol. Indic. 62 (2016) 201--211. https://doi.org/10.1016/j.ecolind.2015.10.030.
\bibitem{doubravsky_dohnal} K. Doubravsk\'{y}, M. Dohnal, Qualitative equationless macroeconomic models as generators of all possible forecasts based on three trend values - Increasing, constant, decreasing, Struct. Change Econ. Dyn. 45 (2018) 30--36. https://doi.org/10.1016/j.strueco.2018.01.001.
\bibitem{dwivedi_sharma_rana_giannakis_goel_dutot} Y. K. Dwivedi, A. Sharma, N. P. Rana, M. Giannakis, P. Goel, V. Dutot, 2023. Evolution of artificial intelligence research in Technological Forecasting and Social Change: Research topics, trends, and future directions. Technol. Forecast. Social Change. 192, 122579. https://doi.org/10.1016/j.techfore.2023.122579.
\bibitem{forbus} K. D. Forbus, Qualitative Modeling, in F. van Harmelen, V. Lifschitz, B. Porter (Eds.), Handbook of Knowledge Representation, Foundations of Artificial Intelligence, chapter, Elsevier, 2008, pp. 361--393.
\bibitem{fu_feng_lande_dmytrenko_manko_prakapovich} M. Fu, J. Feng, D. Lande, O. Dmytrenko, D. Manko, R. Prakapovich, 2021. Dynamic model with super spreaders and lurker users for preferential information propagation analysis. Phys. A 561, 125266. https://doi.org/10.1016/j.physa.2020.125266.
\bibitem{ghosh_das_das22} M. Ghosh, S. Das, P. Das, Dynamics and control of delayed rumor propagation through social networks, J. Appl. Math. Comput. 68 (2022) 3011--3040. https://doi.org/10.1007/s12190-021-01643-5.
\bibitem{ghosh_das_das23} M. Ghosh, P. Das, P. Das, A comparative study of deterministic and stochastic dynamics of rumor propagation model with counter-rumor spreader, Nonlinear Dyn. 111 (2023) 16875--16894. https://doi.org/10.1007/s11071-023-08768-1.
\bibitem{huo_chen_zhao} L. Huo, S. Chen, L. Zhao, 2021. Dynamic analysis of the rumor propagation model with consideration of the wise man and social reinforcement. Phys. A 571, 125828. https://doi.org/10.1016/j.physa.2021.125828.
\bibitem{jahanbakhsh-nagadeh_feizi-derakhshi_ramezani} Z. Jahanbakhsh-Nagadeh, M.-R. Feizi-Derakhshi, M. Ramezani et al., A model to measure the spread power of rumors. J. Ambient Intell. Human Comput. 14 (2023) 13787--13811. https://doi.org/10.1007/s12652-022-04034-1.
\bibitem{jain_walia_gupta} J. Jain, N. Walia, S. Gupta, Evaluation of behavioral biases affecting investment decision making of individual equity investors by fuzzy analytic hierarchy process, Rev. Behavioral Finance 12 (3) (2020) 297--314. https://doi.org/10.1108/RBF-03-2019-0044.
\bibitem{jalali_heidari} M. F. M. Jalali, H. Heidari, 2020. Predicting changes in Bitcoin price using grey system theory. Financ. Innov 6, 13. https://doi.org/10.1186/s40854-020-0174-9.
\bibitem{jamwa_agrawal_sharma_kumar} A. Jamwal, R. Agrawal, M. Sharma, V. Kumar, Review on multi-criteria decision analysis in sustainable manufacturing decision making, Internat. J. Sustain. Eng. 14 (3) (2020) 202--225. https://doi.org/10.1080/19397038.2020.1866708.
\bibitem{jung} C. Jung, Aion: Researches into the Phenomenology of the Self, second revised ed., Princeton University Press, New Jersey, 2014.
\bibitem{konecny_vicha_dohnal} J. Kone\v{c}n\'{y}, T. V\'{i}cha, M. Dohnal, Qualitative phase portrait of modified Black Scholes model. Expert Systems Appl. 37 (5) (2010) 3823--3826. https://doi.org/10.1016/j.eswa.2009.11.037.
\bibitem{kuipers} B. Kuipers, Qualitative Reasoning: Modeling and Simulation with Incomplete Knowledge, MIT Press, Cambridge, Massachusetts, 2003.
\bibitem{lee_kim_jung} H. Lee, J.H. Kim, H. S. Jung, 2025. ESG-KIBERT: A new paradigm in ESG evaluation using NLP and industry-specific customization. Decis. Support Syst. 193, 114440. https://doi.org/10.1016/j.dss.2025.114440.
\bibitem{li_waheed_kirikkaleli_aziz}M. Li, R. Waheed, D. Kirikkaleli, G. Aziz, 2024. Relevance of hybrid artificial intelligence for improving the forecasting accuracy of natural resource prices. Geosci. Front. 15 (3), 101670. https://doi.org/10.1016/j.gsf.2023.101670.
\bibitem{liu_yang_forrest} S. Liu, Y. Yang, J. Forrest, Grey Data Analysis -- Methods, Models and Application, Springer, Singapore, 2017.
\bibitem{mancini_coghill_lusseau}  F. Mancini, G.M. Coghill, D. Lusseau, Using qualitative models to define sustainable management for the commons in data poor conditions, Environ. Sci. Policy 67 (2017) 52--60. https://doi.org/10.1016/j.envsci.2016.11.002.
\bibitem{nwaibeha_chikwendub} E.A. Nwaibeha, C.R. Chikwendub, A deterministic model of the spread of scam rumor and its numerical simulations, Math. Comput. Simulation 207 (2023) 111--129. https://doi.org/10.1016/j.matcom.2022.12.024.
\bibitem{pattanaik_mandal_tripathy} B. Pattanaik, S. Mandal, R.M. Tripathy, A survey on rumor detection and prevention in social media using deep learning, Knowl. Inf. Syst. 65 (2023) 3839--3880. https://doi.org/10.1007/s10115-023-01902-w.
\bibitem{pavlakova-docekalova_doubravsky_dohnal_kocmanova} M. Pavl\'{a}kov\'{a} Do\v{c}ekalov\'{a}, K. Doubravsk\'{y}, M. Dohnal, A. Kocmanov\'{a}, Evaluations of corporate sustainability indicators based on fuzzy similarity graphs, Ecol. Indic. 78 (2017) 108--114. https://doi.org/10.1016/j.ecolind.2017.02.038.
\bibitem{sachdeva_lehal} M. Sachdeva, R. Lehal, Contextual factors influencing investment decision making: a multi group analysis, PSU Res. Rev. 8 (3) 2023 592--608. https://doi.org/10.1108/PRR-08-2022-0125. 
\bibitem{saeed_omlin} W. Saeed, C. Omlin, 2023. Explainable AI (XAI): A systematic meta-survey of current challenges and future opportunities. Knowledge-Based Systems 263, 110273. https://doi.org/10.1016/j.knosys.2023.110273.
\bibitem{salerno_sampagnaro_verdoliva} D. Salerno, G. Sampagnaro, V. Verdoliva, 2022. Fintech and IPO underpricing: An explorative study. Finance Res. Lett. 44, 102071. https://doi.org/10.1016/j.frl.2021.102071.
\bibitem{skapa_bockova_doubravsky_dohnal} S. \v{S}kapa, N. Bo\v{c}kov\'{a}, K. Doubravsk\'{y} and M. Dohnal, Fuzzy confrontations of models of ESG investing versus non-ESG investing based on artificial intelligence algorithms, J. Sustain. Finance Invest. 13 (1) (2023), 763--775. https://doi.org/10.1080/20430795.2022.2030666.
\bibitem{stieglitz_mirbabaie_ross_neuberger} S. Stieglitz, M. Mirbabaie, B. Ross, C. Neuberger, Social media analytics - challenges in topic discovery, data collection, and data preparation. Int. J. Inform. Management 39 (2018) 156--168. https://doi.org/10.1016/j.ijinfomgt.2017.12.002.
\bibitem{sun_sheng_cui} H. Sun, Y. Sheng, Q. Cui, 2021. An uncertain SIR rumor spreading model. Adv. Differ. Equ., 286. https://doi.org/10.1186/s13662-021-03386-w.
\bibitem{verdult_verhaegen} V. Verdult, M. Verhaegen, Filtering and System Identification: A Least Squares Approach, Cambridge University Press, Cambridge, UK, 2007.
\bibitem{wang} X.T. Wang, Financial Decision Making Under Uncertainty: Psychological Coping Methods, in: T. Zaleskiewicz, J. Traczyk (Eds), Psychological Perspectives on Financial Decision Making, Springer, Cham, 2020, pp. 187--201.
\bibitem{wang_song} Q. Wang, P. Song, Is Positive Always Positive? The Effects of Precrisis Media Coverage on Rumor Refutation Effectiveness in Social Media, J. Organizational Comput. Electron. Commerce 25 (1) (2014), 98--116. https://doi.org/10.1080/10919392.2015.990785.
\bibitem{wang_vasilakos_ma_xiong} Y. Wang, A.V. Vasilakos, J. Ma, N. Xiong, On studying the impact of uncertainty on behavior diffusion in social networks, IEEE Trans. Syst. Man Cybern. Syst. 45 (2015) 185--197. https://doi.org/10.1109/TSMC.2014.2359857.
\bibitem{zaslow} E. Zaslow, Quantitative Reasoning: Thinking in Numbers, Cambridge University Press, Cambridge, 2020.
\bibitem{zhang_xie_wang} Q. Zhang, Q. Xie, G. Wang, A survey on rough set theory and its applications. CAAI Trans. Intell. Technol. 1 (4) (2016) 323--333. https://doi.org/10.1016/j.trit.2016.11.001.
\end{thebibliography}
\end{document}